\documentclass[journal, onecolumn]{IEEEtran}

\usepackage{graphicx}
\usepackage{amssymb}
\usepackage{mathrsfs}
\usepackage{caption2}

\ifCLASSINFOpdf
\else
\fi

\begin{document}
%

\title{General Information Theory: Time and Information}
%
%
%

\author{Yilun~Liu,~
        Lidong~Zhu,~\IEEEmembership{Member,~IEEE}
\thanks{Y. Liu and L. Zhu are with the National Key Laboratory of Science and Technology on Communications, University of Electronic Science and Technology of China, Chengdu,
Sichuan, 611731 China e-mail: lyl6205@163.com, zld@uestc.edu.cn.}}

\maketitle
\begin{abstract}
This paper introduces time into information theory, gives a more accurate definition of information, and unifies the information in cognition and Shannon information theory. Specially, we consider time as a measure of information, giving a definition of time, event independence at the time frame, and definition of conditional probability. Further, we propose an analysis method of unified time measure, and find the law of information entropy reduction and increase, which indicates that the second law of thermodynamics is only the law at a certain time measure framework. We propose the concept of negative probability and information black hole to interpret the conservation of information in physics. After the introduction of time, we can give the definition of natural variation and artificial variation from the perspective of information, and point out that it is more reasonable to use the mutation to represent the neural network training process. Further, we point out the defects of the existing artificial intelligence.
\end{abstract}


%
\IEEEpeerreviewmaketitle

\section{Introduction}
%
%
%
%
Information was originally a philosophy. Looking back at the history of information theory, first of all, the literature [1] proposed the definition of information volume. Since then, information has gradually changed from philosophy to scientific research. Later, Shannon proposed the mathematical theory of communication, namely information theory [2], and information changed from science to mathematics, especially information entropy, which belongs to statistics. Further, the three theorems of information theory guide the development of communication and information disciplines, such as coding [3], [4], secure communication [5], [6] and compressed sensing [7]. Shannon information theory theoretically gives the limit of communication performance, so it is regarded as the guiding theory of communication development.

However, aside from communication and other engineering sciences, starting from the information itself, from the first lesson we learned about information theory, the teacher will talk about the difference between Shannon information theory and our cognitive information: one person (Li Hua) tells another person (Xiao Ming) is an important thing, and Xiao Ming can say that the information volume about this matter is zero. Obviously, the information in cognition conflicts with Shannon information theory. So, the first problem raised and solved in this paper is
\begin{itemize}
  \item Can the concept of information in our cognition be unified with Shannon information theory? How to unify?
\end{itemize}

Information is a measure of the event uncertainty. That is, as long as the event has not yet occurred, Shannon information theory is consistent with our cognitive information. After the event occurs, the uncertainty disappears and Shannon information theory is meaningless. Therefore, the information is related to the time of the event. Further, take a more direct example, one person (Li Hua) is in Chengdu, this time on Monday, he is considering whether to go to Beijing on Tuesday or stay in Chengdu, then for Li Hua on Monday, $p(y = Tuesday\;in\;Beijing|x = Monday\;in\;Chengdu) < 1$, that is, the information volume is greater than zero. On Tuesday, Li Hua was indeed in Beijing, so for Li Hua on Tuesday, $p(y|x)=1$, and the information volume is zero. As can be seen from this example, the information volume at different times is different for the same event. Obviously, information is related to time.

In fact, in communication, network information flow [8] can be regarded as a preliminary study of information and time, because the network flow is not negligible in the communication network, that is, the transmission of information is one-way. However, network information theory is not realized, it is because of time. On the other hand, network information theory and Shannon information theory mainly contribute to the coding field, which is an abnormal phenomenon, because signal processing is also a link of communication. Obviously, signal processing is time-dependent, and the description using Shannon information theory is unsatisfactory [9].

The second problem raised and solved in this paper is

\begin{itemize}
  \item How to represent time? How to introduce time into information theory? What is the conflict with the existing information theory after the introduction of time?
\end{itemize}

From the mathematical point of view, there have been studies on measure and probability [10], [11]. Information can be thought of as probability, and time can be seen as a measure of information. It is generally considered that the watershed of probability theory and measure theory is conditional probability and event independence. However, we regard time as a special measure space of probability, which can give the definition of conditional probability and event independence under the framework of time measure, thus avoiding the controversy between probability scholars and measurement scholars. In fact, we can also understand the probability from a functional perspective. Considering $p(x)$ as a function, the value range of $p(x)$ is $[0,1]$, then what is the domain of $p(x)$? The answer given in this paper is time.  It should be noted that the time and random processes in this paper are different. The stochastic process, especially the Markov process, mainly focuses on the study of transition probability.

After the introduction of time, we can do more than just unify the information in Shannon information theory and cognition. Further, we get the law of information changing with time. The third issue raised and solved in this paper is
\begin{itemize}
  \item How does information entropy change over time? How does the information volume change over time?
\end{itemize}

From the perspective of physics, the second law of thermodynamics has described the phenomenon of entropy increase, and also the second law of generalized thermodynamics of black holes [12]. The relationship between thermal entropy and information entropy is comprehensively expounded and verified in [13], [14], [15].

In fact, after the introduction of time, information theory has moved from pure mathematics to physics. The reason why this paper is called general information theory is that the introduction of time information theory is more accurate and versatile than Shannon information theory. In summary, the main contributions of this article are
\begin{itemize}
  \item  Time is a vector, we use frequency and time moment to represent a period of time. Further, we propose the concept of observation time to interpret a cognitive problem: Is the past known? We believe that events are known only when they are within the observation time.
  \item We give a more accurate definition of information volume. Specifically, we define information volume as the event observation time which does not include the moment when the event occurs. Further, after introducing the observation time, we give the definition of event independence and conditional probability.
  \item We find the law of information entropy and information volume at the observation time frame. Specifically, we propose an analysis method, that is, a unified time measure. There are two ways to measure the unified time, different ways will bring different conclusions. In summary, it can be divided into three laws, entropy reduction law, entropy increase law, and information conservation. Our conclusions show that the second law of thermodynamics is only the conclusion of a time measurement framework.
  \item Mathematically, we expand the scope of probability. Specifically, we propose the concept of negative probability and information black hole to improve the shortcomings of current probability theory.
  \item Other contributions, such as we use information to describe neural networks. These contributions are related to specific issues and are not separately stated. The use of general information theory to describe signal processing will be discussed in other paper. This paper is purely theoretical and does not cover specific applications.
\end{itemize}

It should be noted that in this paper, in order to accurately represent the meaning of each word, \emph{information volume} refers to the amount of information. \emph{Time} refers to a period of time, and \emph{time moment} refers to a very short time.

\section{Time and Probability}

\subsection{Vector Representation of Time}
Time is a vector, and the time vector is only one dimension compared to the space vector. Mathematically, time can be represented by only one coordinate axis and the direction of the time axis is from zero to positive infinity. The definition of time moment is given below

\emph{Definition 1:} The \textbf{time moment} is a very short time. The time moment mode is zero and has no direction.

To accurately describe a period of time, at least three elements need to be known: time direction, start time moment and end time moment.
For example, in the time axis, assume ${t_1} > 0,{t_2} > 0$. Define $\overrightarrow {{t_1}{t_2}}$ as a vector from ${t_1}$ to ${t_2}$, with \overrightarrow {{t_1}{t_2}}  =  - \overrightarrow {{t_2}{t_1}}. This representation poses a problem in that it is not certain whether or not this time contains its endpoints, so it is necessary to add the open set and closed set to make the description more accurate. For example, if $\overrightarrow {{t_1}{t_2}}$ contains two endpoints, the exact representation should be written as $\overrightarrow {[{t_1}{t_2}]}$, if only ${t_1}$, it is written as $\overrightarrow {[{t_1}{t_2})}$.

It is worth noting that there is no use of $[{t_1},{t_2})$ simplified $\overrightarrow {[{t_1}{t_2})}$, in order to avoid ambiguity with the following. The following describes another representation of time: frequency.

Another representation for a period of time is frequency, which requires a known frequency, an endpoint. The advantage of this representation is that the direction can be omitted.

For time moment ${t_1},{t_2} \in{\mathbb{R}^ + }$, let the frequency be $F = \frac{1}{{{t_2} - {t_1}}}$, then $[F,{t_1}): = (\overrightarrow {{t_1}{t_2}} ]$, $[\overrightarrow {{t_2}{t_1}} ) =  - (\overrightarrow {{t_1}{t_2}} ] = [ - F,{t_1})$. In this way, we transfer the triple feature of the vector to the modulus of the frequency, the positive and negative of the frequency, and an endpoint. We can judge whether the endpoint is the starting point or the ending point by the positive or negative of the frequency, or compare the length of the time by comparing the modulus of the frequency.

In fact, people are more accustomed to the first representation because the frequency is rarely used. However, the significance of using frequency is that it measures "a period of time" as a certain number on the frequency axis. So that we can compare the length of the two periods by the value of the frequency, and judge the direction of time by the positive and negative of the frequency.

\subsection{Definition of Event Independence}

After the introduction of time, each possible event is tagged with time, so the definition of event independence needs to be corrected, especially the independence of events that may occur at different time. We give the definition of the set after the introduction of time.

\emph{Definition 2:} Each element in the set $X = \{ {x_1},{x_2},...,{x_n}\}$ is an event that may occur, $\sum\nolimits_{i = 1}^n {p({x_i})} {\rm{ = }}1$, $p(x_i)>0$, and each event is mutually exclusive. ${t_0}$ is a time moment, the set $X$ only occurs at ${t_0}$, and must occur at ${t_0}$. This set is recorded as the \textbf{time moment set} ${X_0}: = {X_{{t_0}}}$.

The purpose of the set $X$ occurring only at ${t_0}$ is to avoid confusion on the time axis. If the set occurs over a continuous period of time, it can also be abstracted into a time moment.

According to the definition of event independence, considering the time moment set ${Y_1} = \{ {y_1},{y_2},...,{y_n}\}$, there must be $p({X_0},{Y_1}) = p({X_0})p({Y_1}) = 1$. But if $\exists x \in {X_0}$,  $x \in {Y_1}$, there is an intersection between ${X_0}$ and ${Y_1}$, obviously not independent. Therefore, a more accurate definition of event independence is needed.

\emph{Definition 3:} For the time moment set ${X_0}$ and ${Y_1}$, if there are $p({Y_1} = y|{X_0} = x) = p(y)$, $p({X_0} = x|{Y_1} = y) = p(x)$ for $\forall x \in {X_0}$, $\forall y \in {Y_1}$. Then ${X_0}$ and ${Y_1}$ are independent.

Mathematically, $p({Y_1} = y,{X_0} = x) = p(y)p(x)$ can also represent independence, but it will cause timeline confusion and has no practical meaning. Because if ${t_1} \ne {t_0}$, $x$ and $y$ cannot happen at the same time moment. In fact, we can only stand at one time moment to see another time moment, and it is impossible to concurrently measure things happening at different time moments. For example, we can stand in today to see what happened yesterday, or we can also stand in yesterday to see what happened today, but we cannot be both in yesterday and today, which is the inevitable result of the introduction of time. This phenomenon will be further explained later.
\subsection{Observation Time}

After the introduction of time, there will be a cognitive problem: Is the past known? For example, where were we yesterday? Where were we ten years ago? What happened to the world yesterday, one hundred years ago? Obviously, what has happened is not necessarily known without considering other ways.

In order to avoid this cognitive problem, we propose the concept of observation time ${T_{measure}} = [F,t]$ on the basis of the time axis, and ${T_{measure}}$ is a variable. In this paper , we deem that the time moment of the event belongs to the observation time, the event is known, and for known events, their probability is $1$. Obviously, probability is not a tool for studying known events, then the domain of probability can be represented as the time moment of the event which does not belong to the observation time.

The definition of conditional probability will be discussed in Section III.

\section{General Information Theory}

\subsection{Shannon Information Theory}

For a random event $x$, the definition of information volume is

\begin{equation}
I(x) =  - \log p(x)
\end{equation}

Where $p(x)$ is the probability of event $x$. $X$ is the set of all random events that may occur, and information entropy is defined as
\begin{equation}
H(X) =  - \sum\limits_{x \in X} {p(x)\log p(x)}
\end{equation}

The main contribution of Shannon information theory is that information can be measured. However, Shannon information theory also caused a misunderstanding of information volume: Li Hua told Xiao Ming one thing, Xiao Ming said that information volume on this matter is zero.

As another example, as shown in Fig. 1(a), for the time moment sets ${X_0} = \{ {x_1},{x_2}\}$ and ${Y_1} = \{ {y_1},{y_2}\}$, there are $p({x_1}) = 1 - p({x_2}) = 0.5$, $p({y_1}) = 1 - p({y_2}) = 0.2$, and the transition probability is $\rho {\rm{ = }}\rho {\rm{(}}{y_1}|{x_1}{\rm{) = }}\rho ({x_1}|{y_1}{\rm{)}} = \rho {\rm{(}}{y_2}|{x_2}{\rm{)}} = \rho ({x_2}|{y_2}{\rm{)}} = 1$, where $\rho ({y_j}|{x_i}) = p({y_j}|{x_i})$. According to Shannon information theory, there is $H({X_0}) \ne H({Y_1})$. Now consider the actual situation, as shown in Fig. 1(b). Without loss of generality, let ${t_0} < {t_1}$, now consider $p({x_1}) = 0.5$ and $\rho {\rm{(}}{y_1}|{x_1}{\rm{) = 1}}$, then $p({y_1}) = \rho {\rm{(}}{y_1}|{x_1}{\rm{)}}p({x_1}) = 0.5$, and $p({y_2}) = 0.5$, obviously $H({X_0}) = H({Y_1})$. Looking back at these two results, regardless of time, the hypothesis of Fig. 1(a) is no problem, but in reality this will not happen.

\begin{figure}[!ht]
  \centering
  \includegraphics[width=4.0in]{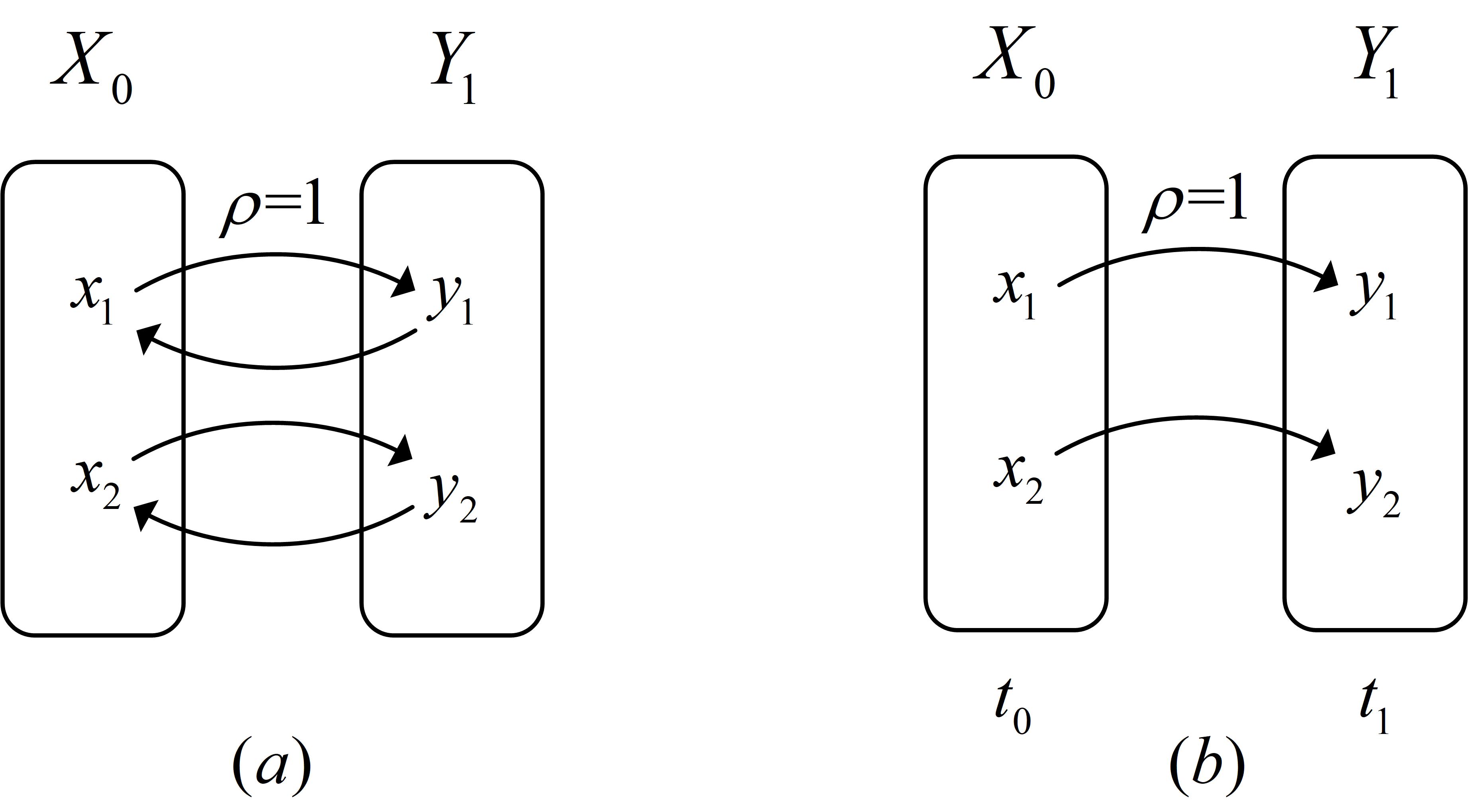}\\
  \captionsetup{justification=centering}
  \caption{(a) Pure mathematical model, (b) Time mathematical model.}
\end{figure}

Both of the above models can be analyzed by Shannon information theory, because it does not take into account the time factor, and many scenes that cannot happen in reality can be analyzed. However, these scenes only exist in mathematical theory, which are easy to become "counter-examples." The key idea of this paper is also about the relationship between time and information, and proposes a general information theory.

\subsection{The Proposed Information Theory}

\subsubsection{Information and Knowledge}
After introducing the concept of observation time, the definition of information volume can be further refined

\emph{Definition 3:} For time moment set ${X_0}$, event $x$ occurs at time moment $t_0$, $x \in X$ and $F \in \mathbb{R}$. The \textbf{information volume} of $x$ is
\begin{equation}
I(x) =  - \sum\limits_{x \in {X_0}} {\log p(x)} \;\;\;\;{t_0} \notin {T_{p(x)}} = [F,t]
\end{equation}

By \emph{definition 3}, the information entropy of the time moment set ${X_0}$ can be rewritten as
\begin{equation}
H({X_0}) =  - \sum\limits_{x \in {X_0}} {p(x)\log p(x)} \;\;\;\;{t_0} \notin {T_{p(x)}} = [F,t]
\end{equation}

In (3), ${T_{p(x)}}$ is the observation time of ${p(x)}$, which is not specified in Shannon information theory. According to the definition, the necessary and sufficient condition for information volume to be zero is that there is only one element in ${X_0}$, that is, ${p(x)}=1$.

So, what is $x$ when ${t_0} \in {T_{p(x)}}$? The definition of knowledge is

\emph{Definition 4:} For the time moment set ${X_0}$, event $x$ occurs at time moment $t_0$, $x \in X$ and $F \in \mathbb{R}$. If ${t_0} \in {T_{k(x)}} = [F,t]$, then $x$ is \textbf{knowledge}, denoted as $K(x)$.

Knowledge is a known event that has occurred, is an intrinsic property, and the measure of knowledge is artificially determined. Knowledge and information are two different attributes of things, and the information volume of knowledge is zero.

The observation time of knowledge and information is different, and the difference is whether or not ${t_0}$ is included. It is very interesting that we find that the introduction of time will cause the problem consistent with the face of quantum mechanics. This problem can be described as follow
\begin{itemize}
  \item If the observation is not made at time moment ${t_0}$, how do we know that $x$ is occurring?
\end{itemize}

This is consistent with the question of Schrodinger's cat. There are many discussions on quantum mechanics, such as the Copenhagen Interpretation, Einstein hidden variable theory, multi-world interpretation, and so on. This paper is not intended to explain this issue, the following is a discussion based on the theory proposed in this paper.

The observation time ${T_{k(x)}}$ of $K(x)$ contains ${t_0}$, while the ${T_{p(k(x))}}$ of $I(K(x))$ does not contain ${t_0}$, because ${T_{k(x)}}$ and ${T_{p(k(x))}}$ are two observation time, which is equivalent to two observations. The order of observation is obviously from ${T_{k(x)}}$ to ${T_{p(k(x))}}$, and $K(x)$ appears with $p(K(x)) = 1$ after ${t_0}$ on the time axis, and $x$ is meaningless before ${t_0}$, then $I(k(x))=0$ is valid. We use the \emph{step effect of knowledge} to describe the curve of the probability of $k(x)$ to over time, as shown in Fig. 2. This is consistent with the increasing knowledge of human society.
\begin{figure}[!ht]
  \centering
  \includegraphics[width=1.5in]{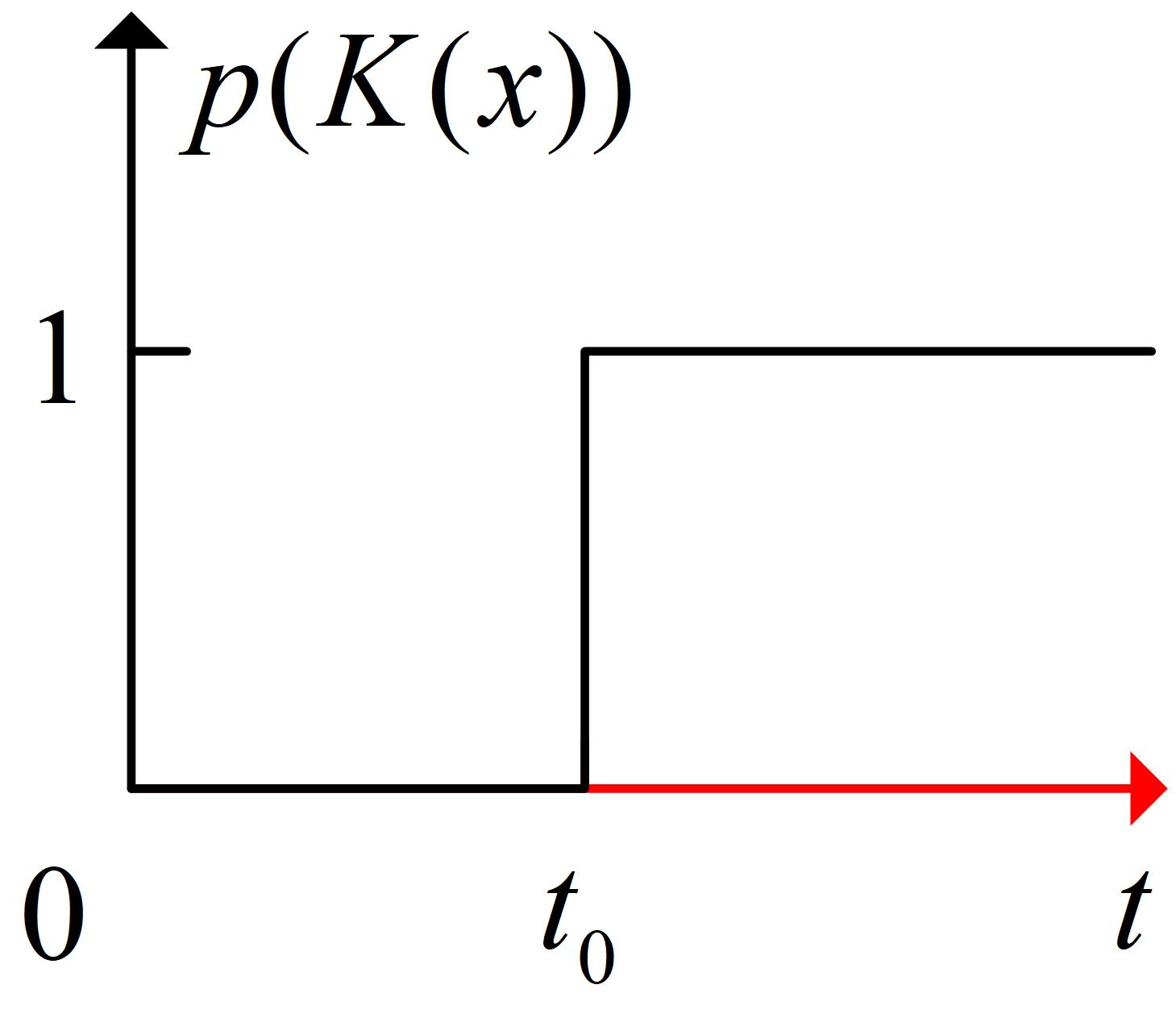}\\
  \captionsetup{justification=centering}
  \caption{The step effect of knowledge}
\end{figure}

In summary, $K(x)$ denotes the knowledge contained in $x$, which is a intrinsic property of $x$. The information volume of $K(x)$ is zero. $I(x)$ represents the amount of information of $x$, which is a measure of the uncertainty of $x$. The correspondence between the wave function collapse effect of quantum mechanics in information theory is the step effect of knowledge.

\subsubsection{Mutual Information and Conditional Information}
Considering another time moment set ${Y_1}$, mutual information volume $I(x,y)$ is defined as follow

\begin{equation}
I(x,y) = -\sum\limits_{x,y} {p(x,y)\log \frac{{p(x,y)}}{{p(x)p(y)}}} \;{t_0},{t_1} \notin {T_{x,y}}
\end{equation}

In (5), there are $x \in {X_0}$,$y \in {Y_1}$, and ${T_{x,y}}:={T_{p(x,y)}}$. In fact, the mutual information volume exists only in mathematical theory, which is due to the problem after the introduction of time, and the explanation is given below. Joint distribution probability is

\begin{equation}
p(x,y) = p(y|x)p(x)
\end{equation}

It can be known from (6) that the joint distribution probability can be expressed as the product of a certain event probability and a conditional probability. For $p(y|x)$, there is ${t_0} \in {T_{y|x}}$, and for $p(x)$, ${t_0} \notin {T_{y|x}}$, which is obviously contradictory. The only reasonable explanation is that two observations were made during the acquisition of $p(x,y)$, the first observation is $p(y|x)$ and the second observation is $p(x)$. However, the definition of mutual information volume $I(x,y)$ only contained one observation. Therefore, mutual information is a mathematical hypothesis that does not exist in reality.

The above interpretation has a cognitive problem: Can two observations are performed at the same time moment? In theory, it can be considered that observations can be made simultaneously. In fact, the difficulty of achieving two simultaneous observations is equivalent to the brain of the person thinking about two things simultaneously. Therefore, in this paper, mutual information volume is used only as an analytical tool for formula transformation.

For the condition information, when ${t_0} = {t_1}$ or ${X_0}$, ${Y_1}$ are independent, condition information entropy $H({Y_1}|{X_0})$ is

\begin{equation}
H({Y_1}|{X_0}) =  - \sum\limits_{x \in {X_0}} {p(x)\sum\limits_{y \in {Y_1}} {p(y|x)\log p(y|x)}}
\end{equation}

There is no observation time in (7) because two observations are made, ${T_x}$ and ${T_{y|x}}$ respectively.

If ${t_0} < {t_1}$, the value of the conditional information entropy will depend on the relationship between ${X_0}$ and ${Y_1}$, as well as the problem of determining the range of observation time. In fact, it is difficult to judge whether two events are related, and many things that seem to be unrelated are also related in time. For example, the "butterfly effect", a butterfly in the Amazon rainforest agitated a wing a month ago, causing a hurricane on the Atlantic. Therefore, conditional information entropy can only be analyzed when faced with specific problems.

Considering the mutual information formula as follow
\begin{equation}
I({X_0},{Y_1}) = H({X_0}) - H({X_0}|{Y_1}) = H({Y_1}) - H({Y_1}|{X_0})
\end{equation}

The above equation is established in the field of probability theory, but after the introduction of time, ${T_{x,y}} \ne {T_{x|y}} \ne {T_{y|x}}$. From the perspective of the set, there is an intersection of these observation times, but from the perspective of time measure, the constraints of the three are mutually exclusive. Therefore, mutual information, information entropy and conditional information entropy cannot be in an equation at the time measure, which is also the unsolved problem of measure theory and probability theory.

In summary, Shannon information theory can be a special case of information theory proposed in this paper at ${t_0}={t_1}$, and coding can be regarded as a Markov process with a transition probability of 1. Therefore, the pre-encoding time moment and post-encoding time moment can be abstracted to the same time moment.
\subsection{Unified Time Measure}

In fact, after the introduction of the observation time, we have not strictly imposed it, only used it to represent the domain of probability. However, in practice, we will pay attention to the change of information at different time moment, and (8) is meaningless if it does not constrain the observation time. In general, we pay attention to the change of information entropy ${H(X_0)}$, ${H(Y_1)}$ at different time moments, so we propose a method for setting the observation time for analyzing the change of information entropy

\textbf{Criterion:} Set an endpoint of the observation time at the time moment that an event occurs on the time axis, if the conditional probability does not require other time moment to be known, it is unknown.

According to the above-mentioned method and \emph{definition 4}, firstly, it is judged whether the event is knowledge in the unified time measure, and if it is knowledge, the information volume and the information entropy are both zero. If not, the information entropy is consistent with the definition.

For example, there are two time moment set ${X_0} = \{ {x_1}\}$, ${Y_1} = \{ {y_1},{y_2}\}$ and ${\rho _{11}} = \rho ({y_1}|{x_1}) = 1 - \rho ({y_2}|{x_1}) = 1 - {\rho _{12}} = \alpha $,  $0 < \alpha  < 1$, as shown in Fig. 3(a). Let the observation time be: ${T_m} = (F,{t_1}]$, and the event ${y_1}$ occurs at the time moment ${t_1}$. Then, our analysis method has the transition probability of 1 when standing at ${t_1}$. This example is shown in Fig. 3(b).

\begin{figure}[!ht]
  \centering
  \includegraphics[width=5.0in]{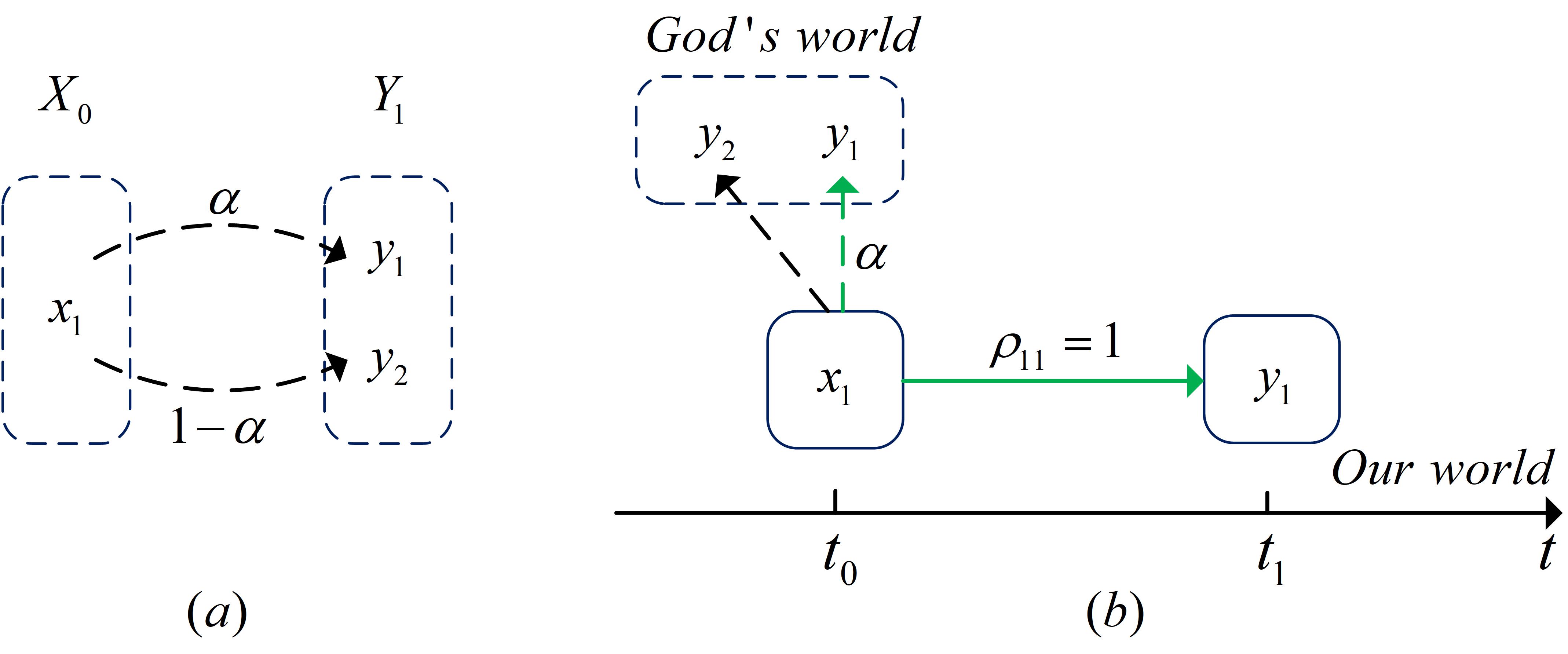}\\
  \captionsetup{justification=centering}
  \caption{(a) Stochastic process model (b) Unified time measure model.}
\end{figure}

In fact, if there are multiple elements in ${X_0}$, it needs to be classified and discussed, and will not be described here. Fig. 3(b) illustrates the difference between our method and the traditional method: the traditional method is God's perspective, and it is at the earlier time moment to see what has happened, and our method is to see what has happened at the later time moment. If what happened in the past is known, then the current result is inevitable.

The unified time measure ${T_m} = (F,{t_1}]$ can also be interpreted with probability. For the prior probability $p(y|x)$, $y$ is known, so $p(y|x)=1$. For the posterior probability, the analysis method is consistent with the probability theory, because our observation time does not require the inclusion of ${t_0}$. Similarly, if the uniform time measure is set to ${T_m} = (F,{t_0}]$, then the conclusion is the opposite. The law of entropy increase and reduction are the conclusions at different unified time measures.

We can also use popular vocabulary to describe the prior probability and posterior probability in general information theory with ${T_m} = (F,{t_1}]$. For example, memory (prior probability) is deterministic, while recall (posterior probability) is uncertain. For example, Xiao Ming knows what he is doing today a year ago?

\subsection{Information Entropy and Time}

\subsubsection{Special Time Moment Set and Lemma 1}
First, because the transition probability of time is 1 and the transition probability between different elements is not necessarily 1, we use the \emph{operator variable} $O$ to represent the possible processing between $t_0$ and $t_1$. The conclusion reached is Lemma 1

\emph{Lemma 1:} For the time moment sets ${X_0}$, ${Y_1}$, ${t_1} > {t_0}$, there is a mapping $M:{X_0} \to {Y_1}$. And for $\forall x \in {X_0}$, $\exists y \in {Y_1}$, there is an operator variable $O:x \to y$, then $M$ is surjection (onto mapping).

\emph{proof:} It is easy to know that $\forall x \in {X_0}$, $\exists y \in {Y_1}$ is connected to $x$, and only need to prove that $\forall y \in {Y_1}$, $\exists x \in {X_0}$ is connected to $y$.

Assuming that $\exists {y_n} \in {Y_1}$ and $\forall x \in {X_0}$ is not connected to ${y_n}$, it is equivalent to ${X_0}$ is transferred to ${Y_1}-{y_n}$ with transition probability ${\rho _M}=1$, there is $p({Y_1} - {y_n}) = {\rho _M}p({X_0}) = 1$. From the definition of the time moment set, there is $p({Y_1} - {y_n}) < 1$ because of $p({y_n}) > 0$, $p({Y_1}) = 1$. This is contradictory.

Therefore, $M$ is surjection.

\subsubsection{Entropy Reduction Law} As mentioned above, the law of change in entropy is related to the unified time measure. Theorem 1 is the law of entropy reduction, as shown below

\emph{Theorem 1:} For the time moment sets ${X_0}$, ${Y_1}$, ${t_1} > {t_0}$, there is a mapping $M:{X_0} \to {Y_1}$. And for $\forall x \in {X_0}$, $\exists y \in {Y_1}$, there is an operator variable $O:x \to y$. If unified time measure is ${T_m} = (F,{t_1}]$, then $H({Y_1}) \le H({X_0})$.

\emph{proof:} We introduce the observation time ${T_m} = (F,{t_1}]$ into (8), as shown below

\begin{equation}
H({X_0}) - H({X_0}|{Y_1}) = H({Y_1}) - H({Y_1}|{X_0}), {T_m} = (F,{t_1}]
\end{equation}

Due to the limitation of the observation time ${T_m}$, ${H(Y_1)}={H(K(Y_1))}=0$, By \emph{Lemma 1}, $M$ is onto mapping, that means for $\forall y \in {Y_1}$, there is $\rho (y|\exists x \in {X_0}) = 1$ and then $H({Y_1}|{X_0}) = 0$. So, we can get

\begin{equation}
H({X_0}) = H({Y_1}) + H({X_0}|{Y_1})
\end{equation}

And because of $H({X_0}|{Y_1}) \ge 0$, so $H({Y_1}) \le H({X_0})$.

\emph{Inference 1.1:} The necessary and sufficient conditions of $H({Y_1}) = H({X_0})$ is that the mapping $M$ is a one-to-one mapping.

\emph{proof:} Sufficiency. If the mapping $M$ is a one-to-one mapping, then $p(\exists x \in {X_0}|\forall y \in {Y_1}) = 1$, and there are $H({X_0}|{Y_1}) = 0$, $H({Y_1}) = H({X_0})$.

Necessity. If the mapping $M$ is not a one-to-one mapping, then $p(\exists x \in {X_0}|\exists y \in {Y_1}) < 1$, and there are $H({X_0}|{Y_1}) > 0$, $H({Y_1}) < H({X_0})$, which conflicts with known. Therefore, the mapping $M$ is a one-to-one mapping.

\emph{Inference 1.2:} $H({X_0}) = H({X_0}|{Y_1})$.

\emph{proof:} Because $H({Y_1})=0$, then $H({X_0}) = H({X_0}|{Y_1})$.

\emph{Entropy reduction law} and \emph{inference 1.1} will play an important role in signal processing and theoretical analysis of signal processing using neural networks, but this paper does not discuss specific issues. The relevant issues will be discussed in the next paper.
\emph{inference 1.2} can be understood that the information entropy of the past time moment is equal to the information entropy of the recall. This also indicates that the past is not necessarily known.

\subsubsection{Entropy Increase Law} The second low of entropy is the same as the second law of thermodynamics, as shown below

\emph{Theorem 2:} For the time moment sets ${X_0}$, ${Y_1}$, ${t_1} > {t_0}$, there is a mapping $M:{X_0} \to {Y_1}$. And for $\forall x \in {X_0}$, $\exists y \in {Y_1}$, there is an operator variable $O:x \to y$. If unified time measure is ${T_m} = (F,{t_0}]$, then $H({Y_1}) \ge H({X_0})$.

\emph{Inference 2.1:} The necessary and sufficient conditions of $H({Y_1}) = H({X_0})$ is that the mapping $M$ is a one-to-one mapping.

\emph{Inference 2.2:} $H({Y_1}) = H({Y_1}|{X_0})$.

The proof process is consistent with \emph{Theorem 1}, \emph{Inference 1.1}, and \emph{Inference 1.2}.

Entropy in thermodynamics can be used to describe the degree of material chaos. For example, by dropping ink into water, the degree of "chaos" of ink and water increases, so it is entropic. In fact, the increase in entropy in thermodynamics is only for "ink" or "water". Taking ink as an example, if the ink is not dripped into the water, its state is only a state of "ink", and "dropping into water" is equivalent to an unknown operator variable, resulting in a state other than "ink", so the entropy increases. However, if we treat ink and water as a whole, their state will always change within a known set of states, that is, entropy does not increase.

In essence, the increase or decrease of entropy depends on whether the operator variable is known. If the operator variable is known, it will not increase entropy. If it is unknown, it will increase entropy. That is to say, the second law of thermodynamics is equivalent to predicting the law of information entropy change from now on, while \emph{theorem 1 is} the law of change of information entropy from the past to the present.

\newcommand{\RNum}[1]{\uppercase\expandafter{\romannumeral #1\relax}}
The law of change in information volume will be discussed in the Section \RNum{4}.

\section{Information and Physics}

\subsection{Negative Probability, Information Black Hole and Information Volume Conservation}

If the observation time is different, the law of entropy change is different, as described in \emph{Theorem 2} and \emph{Theorem 1}. However, in the second law of thermodynamics, the future state is unknown, so we only consider the change of information volume in the framework of \emph{Theorem 1}.

As shown in Fig. 3(b), the transition probability $\rho ({y_1}|{x_1}) = 1$. Regardless of why ${y_1}$ occurs, there will still be a problem: Where is the other state ${y_2}$? This is also one of the debates in quantum mechanics. We think about this from the perspective of information volume.

Section \RNum{2} shows the limitations of the existing probability theory after introducing the time measure. Let us further improve the probability theory. In mathematics, the complex variable has $\sqrt { - 1}  = i$, and the imaginary number is used to represent the number that does not exist in the real field. We extend the range of probability to improve the existing probability.

\emph{Definition 5:} For the time moment sets ${X_0} = \{ {x_1}\}$, ${Y_1} = \{ {y_1},{y_2}\}$, there is ${\rho _{11}} = \rho ({y_1}|{x_1}) = 1 - \rho ({y_2}|{x_1}) = 1 - {\rho _{12}} = \alpha $,  $0 < \alpha  < 1$, and the event ${y_1}$ occurs at the time moment ${t_1}$. Then, the transition probability ${\rho _i}$ of the event ${y_2}$ that may occur but does not occur is \textbf{negative probability}. And there is ${\rho _i} =  - \rho ({y_2}|{x_1}) = \alpha  - 1$.

In the negative probability framework, we can think that $y_2$ has occurred, but it exists in a state that can no longer be transferred after entering, which is called \textbf{information black hole}. Obviously, information black hole is time-independent, and information black hole exists at every time moment. Thus, Fig. 3(b) will become as shown in Fig. 4.

\begin{figure}[!ht]
  \centering
  \includegraphics[width=5.0in]{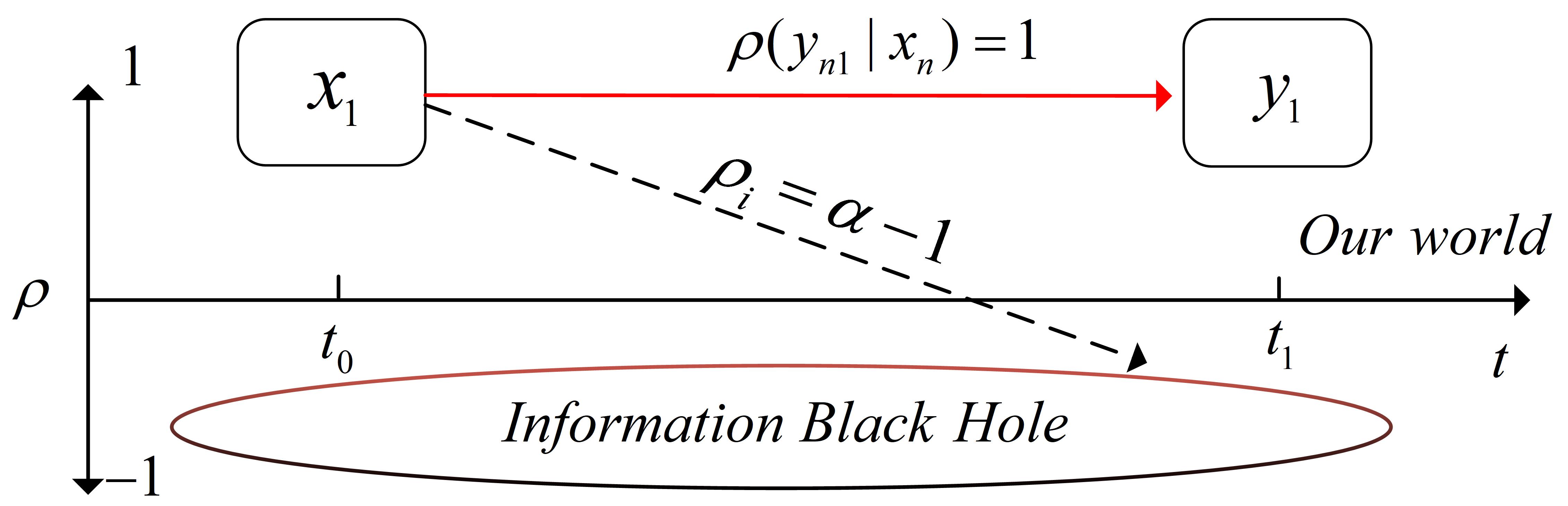}\\
  \captionsetup{justification=centering}
  \caption{Information black hole and negative probability.}
\end{figure}

In the information black hole and negative probability framework, it is easy to obtain the conservation of information volume.

\emph{Theorem 3:} For the time moment sets ${X_0} = \{ {x_1}\}$, ${Y_1} = \{ {y_1},{y_2}\}$, there is ${\rho _{11}} = \rho ({y_1}|{x_1}) = 1 - \rho ({y_2}|{x_1}) = 1 - {\rho _{12}} = \alpha $,  $0 < \alpha  < 1$, and the event ${y_1}$ occurs at the time moment ${t_1}$. Then, ${I(x_1)}={I(y_1)}$.

\emph{proof:} $I({y_1}) =  - \log (\rho ({y_1}|{x_1})p({x_1})) = I({x_1})$.

In fact, \emph{Theorem 3} describes the conservation of information volume rather than the conservation of information entropy, because entropy is a statistical concept, not a real event. On the other hand, if the time measure framework of \emph{Theorem 1} is not used, the information volume is not conserved using the framework of \emph{Theorem 2}. Both are reasonable.

In summary, in the negative probability framework, the multi-world interpretation is no longer established, and the state that has not occurred enters the information black hole.
\subsection{Information and Mechanics}

The advantage of information theory is that it describes the world from a probabilistic perspective, and when a certain phenomena occurs with the probability of 1, we often give it a new meaning. For example, many natural phenomena are events with the probability of 1, for example, Newton observed that apples falling from trees must fall to the ground, satellites orbiting the Earth in space, and so on. In physics, gravity is used to describe this natural phenomenon and is promoted.

Obviously, from the perspective of information, gravity is only the physical abstraction of the inevitable event. If we use mechanics to describe accidental events such as quantum mechanics, the macroscopic performance is not accurate.

From the perspective of the four fundamental interactions, Einstein unified field theory has not yet been completed. Information theory and mechanics are not directly related, so using information to describe accidents in physics is a feasible path.
\subsection{Relativity of Information}
The key point of this paper is that information is time-related. From a physics perspective, information is also spatially related. Back to the example of Li Hua who goes to Beijing on Tuesday. For Xiao Ming, without regard to other means, even on Tuesday, Xiao Ming did not know whether Li Hua was in Chengdu or not, and ${p_{XiaoMing}}(y|x) < 1$. Therefore, information is relative.

However, we usually do not pay attention to Xiao Ming's information, so the relative nature of information exists only as a physical property. In essence, the reason why information is relative is that the observation time caused by spatial position is relative, which is consistent with the special theory of relativity [16].

\section{Information and Neural Networks}

\subsection{Natural Variation and Artificial Variation}

Generally, we believe that intelligence is generated by variation and evolution. The following are definitions of natural variation and artificial variation.

\emph{Definition 6:} For the time moment sets ${X_0} = \{ {x_1}\}$, ${Y_1} = \{ {y_1},{y_2}\}$, there is ${\rho _{11}} = \rho ({y_1}|{x_1}) = 1 - \rho ({y_2}|{x_1}) = 1 - {\rho _{12}} = \alpha $,  $1 \gg \alpha  \to 0$, and the event ${y_1}$ occurs at the time moment ${t_1}$, calling this process \textbf{natural variation}.

\emph{Definition 7:} For the time moment sets ${X_0} = \{ {x_1}\}$, ${Y_1} = \{ {y_1},{y_2}\}$, there is ${\rho _{11}} = \rho ({y_1}|{x_1}) = 1 - \rho ({y_2}|{x_1}) = 1 - {\rho _{12}} = 1 $, and the event ${y_2}$ occurs at the time moment ${t_1}$, calling this process \textbf{artificial variation}.

Natural variation indicates that an event with a very low probability has occurred and usually takes a long time. Artificial variation is the event that "cannot happen in a short time." In essence, artificial variation is also a natural variation and refers to the occurrence of natural variation that cannot occur in a short period of time by some methods. For example, ultraviolet radiation illuminates plant seeds to achieve variation.

\subsection{Artificial Variant Neural Networks}

From the perspective of the information volume and the operator variable $O$, there is $I({y_2}) = I(O({x_1})) = \infty $. That is to say, the artificial variation is unexplained when the operator operator variable $O$ is known, because the operator variable $O$ is subject to "artificial interference". The conceptual diagram of the artificial variant neural network is shown in Fig. 5.
\begin{figure}[!ht]
  \centering
  \includegraphics[width=5.0in]{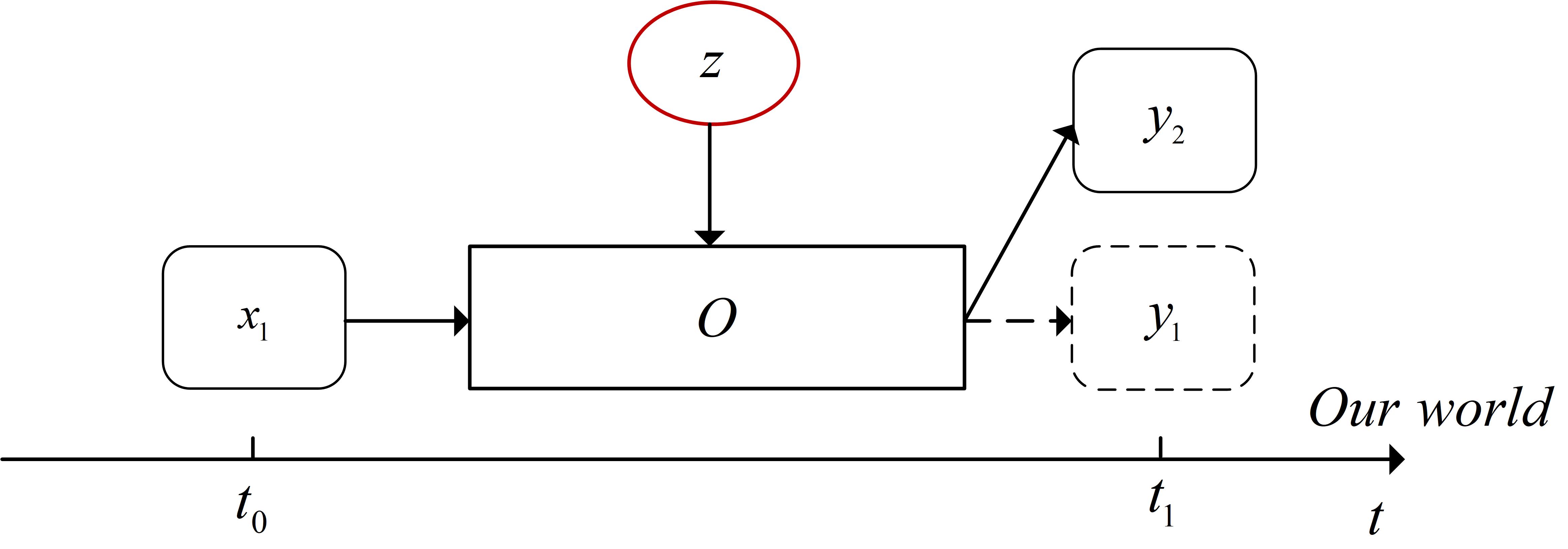}\\
  \captionsetup{justification=centering}
  \caption{Artificial variant neural network diagram.}
\end{figure}

Variable automatic encoder [17] is an artificial variant neural network. Obviously, if there is a suitable "artificial interference", it will help to accelerate the training of the neural network.

Considering evolution, evolution is usually the result of natural selection or manual selection. Whether it is artificial selection or natural selection, the goal of evolution is ¡°survival of the fittest¡±. But the degree of rigor of artificial and natural selection is different. Generative adversarial network [18] is a neural network that artificially mutates and performs natural selection.

\subsection{Artificial Intelligence?}

The neural network is an architecture, and "learning" is more for the already intelligent organisms. The "variation" and "evolution" describe the training process of the neural network, which seems more appropriate than "learning". In view of the functions of the existing neural networks, we think that the current artificial intelligence has developed into the stage of ¡°animal intelligence¡±, that is, it can realize the tasks that humans have given to them, and even have spatial imagination [19]. The difference between "humanoid intelligence" and "animal intelligence" is whether it has independent consciousness and thinking, and can complete innovation without human intervention. We present a block diagram of the hypothesis, as shown in Fig. 6.

\begin{figure}[!ht]
  \centering
  \includegraphics[width=4.0in]{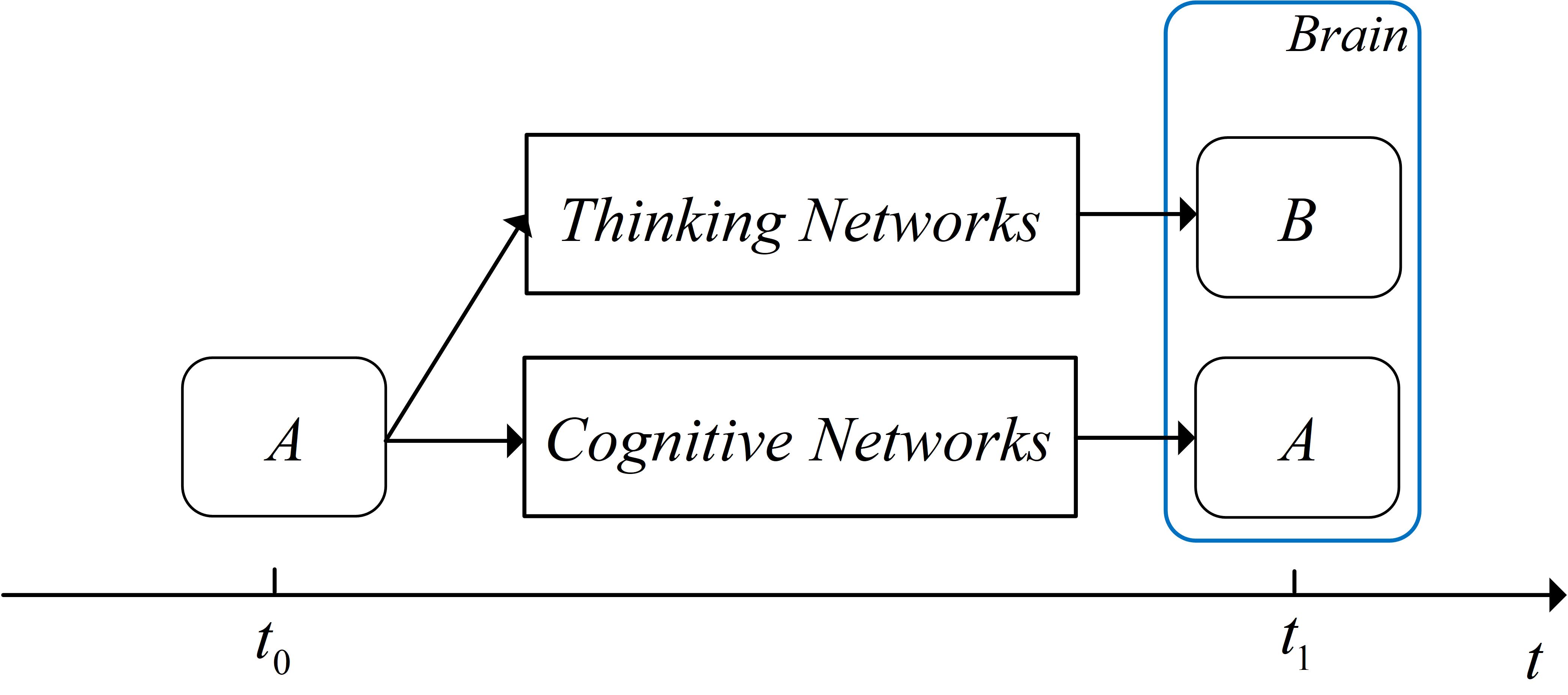}\\
  \captionsetup{justification=centering}
  \caption{An intelligent form of implementation.}
\end{figure}

In Fig. 6, $A$ refers to the observed things, and the training process of the thinking network and the cognitive network is independent. The cognitive neural network passes the observed $A$ to the brain, and the thinking network realizes innovation. The thinking network is usually not activated, it may be activated by means of memory, and the variable operator $O$ in the thinking network seems to be difficult to train and in the cognitive neural network it is easy to train.
\section{Conclusion}
In this paper, we propose a general information theory. Specifically, we introduce time into the representation of information volume and information entropy, giving a more accurate definition of information.

We propose the concept of observation time, thinking that only events in the observation time are known, and the past is not necessarily known. We propose a unified time measure method for general information theory, and further discover the entropy increase and decrease law.

We propose the concept of information black hole and negative probability, in which the information volume is conserved. If we regard $x$ as a quantum, the conservation of information volume can link the most macroscopic existence of the universe with the most microscopic particles. This is also our windfall.

Further, we believe that the use of mutation and evolution to describe the neural network training process is more appropriate. For the development of artificial intelligence, we believe that it is still in the stage of ¡°animal intelligence¡±. To realize human-like intelligence, further research on the human brain is needed.

In summary, this paper studies information in our cognition, not information in mathematics. In essence, mathematics and physics are tools used to interpret the laws of the world, and the only criterion for testing mathematics or physics is practice. Therefore, we will also use general information theory to interpret neural networks and its application in signal processing in our next paper [20].

Finally, information black hole is beyond time, and the great thing about human beings is that human civilizations, such as family, knowledge, architecture, and so on, have also surpassed time.


%

\section*{Acknowledgment}

The major of the authors is not physics and mathematics. In writing this paper, the physical content is also the inevitable result after the introduction of time. It may not be comprehensive enough to refer to the physical related references, and we would like to express our gratitude.

In the process of writing this paper, many selfless netizens in the knowledge sharing platform allow us to quickly and accurately understand the progress in physics and mathematics, and we would like to express our gratitude.

This work is funded by the National Natural Science Foundation (No.61871422) and the Ministerial Foundation (No.61405180\\50316).
\ifCLASSOPTIONcaptionsoff
  \newpage
\fi



%

%

\begin{IEEEbiographynophoto}{Yilun Liu}
received the B.S. degree in communication engineering from University of Electronic Science and Technology of China, Chengdu,
 China, in 2018. Since Sept. 2018, he is studying for the M.S. degree in University of Electronic Science and Technology of China.
His research interests include information theory, communications, and signal processing.
\end{IEEEbiographynophoto}

\begin{IEEEbiographynophoto}{Lidong Zhu}
(M'07) is a full professor of University of Electronic Science and Technology of China. He received the B.S. and M.S. degree from Sichuan University in 1990 and 1999 separately. In 2003 he got the Ph.D. degree in University of Electronic Science and Technology of China.
His interest includes satellite communications, communication signal processing, secure communications.
\end{IEEEbiographynophoto}




\end{document}